\documentclass[twoside,twocolumn,aps,prb,superscriptaddress,longbibliography]{revtex4-2}
\usepackage[latin9]{inputenc}
\usepackage{array}
\usepackage{mathrsfs}
\usepackage{multirow}
\usepackage{amsmath}
\usepackage{amssymb}
\usepackage{graphicx}
\PassOptionsToPackage{version=3}{mhchem}
\usepackage{mhchem}

\makeatletter

\providecommand{\tabularnewline}{\\}

\usepackage{float}
\usepackage{soul}
\usepackage{changes}
\usepackage{mathrsfs}
\usepackage{comment}
\usepackage{multirow}
\usepackage{xcolor}

\makeatother

\begin{document}
\title{Predicting and understanding diffusion lengths and lifetimes in solids via a many-body \textit{ab initio} method: The role of coupled dynamics}
\setcounter{page}{1}
\date{\today}
\author{Junqing Xu}
\email{jqxu@hfut.edu.cn}
\affiliation{Department of Physics, Hefei University of Technology, Hefei, Anhui, China}
\begin{abstract}
We present an \textit{ab initio} method of diffusion, relaxation
and dephasing processes of arbitrary observables, and corresponding
diffusion lengths and lifetimes in solids. The method is based on
linearized density-matrix master equation, with quantum treatment
of electron scattering processes. It enables clear \textit{ab initio}
descriptions of long lifetimes and diffusion lengths using approximate
formulas at different levels, such as Dyakonov-Perel and drift-diffusion
relations for spin decay and those beyond with coupled dynamics. Our
results of graphene-hBN show that the coupling between dynamical processes
can significantly affect spin diffusion and relaxation. Our method
provides a transparent and powerful tool for predicting and understanding
diffusion and relaxation.
\end{abstract}
\maketitle
Diffusion lengths $l$ and lifetimes $\tau$ of observables such as
spin, pseudo spin besides carrier occupation, of bulk or itinerant
electrons are key parameters in spintronics, electronics, etc., and
are critical to developing next generation low-power electronics.\citep{vzutic2004spintronics,avsar2020colloquium,sierra2021van,sohn2024dyakonov}
\textit{Ab initio} studies of them are invaluable to searching new
materials and understanding decay mechanisms, but were only carried
out for limited materials.\citep{xu2024spin,park2022predicting,restrepo2012full,xu2024graphite,nair2021spin,belashchenko2016theory}

For spin diffusion length $l_{s}$, fully \textit{ab initio} simulations,
beyond drift-diffusion (DD) relation and model Hamiltonians\citep{wu2010spin,vila2021low},
were only done recently for a small number of materials\citep{nair2021spin,belashchenko2016theory,rojas2019ab,egami2021calculation}.
However, in their works, phonons are either missing or considered
through random atomic movements. Thus, quantum treatment of the electron-phonon
(e-ph) interaction remains difficult, limiting predictive accuracy.
For spin lifetimes $\tau_{s}$, \textit{ab initio} methods were developed
by several groups since 2012.\citep{xu2024spin,park2022predicting,restrepo2012full}
Among these, we recently developed a method of spin relaxation, dephasing
and $\tau_{s}$,\citep{xu2024spin,xu2023ab,sundararaman2017jdftx}
based on real-time evolution of density-matrix (DM) master equation
(ME) with quantum treatment of electron scattering processes. The
method was applied to disparate materials and is in principle applicable
to $\tau$ of arbitrary observables. It however has a few limitations:
It cannot systematically derive approximate formulas of long $\tau$;
It is difficult to interpret complex decay processes when observable
evolution curve is complicated.

\begin{table*}
\begin{tabular}{c|c|c}
\hline 
\hline 
Approx. & $V^{KR}$ & Formula of $\Gamma_{s}$ or $\lambda_{s}$\tabularnewline
\hline 
\hline 
EY & $\left\{ \rho^{s_{i}}\right\} $ & $\Gamma_{s,i}^{\mathrm{EY}}=-\varrho^{s_{i},\dagger}L\rho^{s_{i}}=-\varrho^{s_{i},\dagger}L^{C}\rho^{s_{i}}$\tabularnewline
\hline 
\multirow{2}{*}{E+D} & \multirow{2}{*}{$\left\{ \rho^{s_{i}},-L\rho^{s_{i}}\right\} $} & $\Gamma_{s,i}^{\mathrm{E+D}}\approx\frac{1}{2}(\Gamma_{s,i}^{\mathrm{EY}}+\tau_{p}^{-1})-\frac{1}{2}[(\tau_{p}^{-1}-\Gamma_{s,i}^{\mathrm{EY}})^{2}-4\overline{\Omega_{\perp i}^{2}}]^{1/2},$\tabularnewline
 &  & $\overline{\Omega_{\perp i}^{2}}=-\varrho^{s_{i},\dagger}\left(L^{e}\right)^{2}\rho^{s_{i}}$\tabularnewline
\hline 
RR-2, ${\bf B}||\alpha$ & $\{U_{s_{\beta}}^{0R},U_{s_{\gamma}}^{0R}\}$, $\beta\neq\gamma\perp{\bf B}$ & $\Gamma_{s}\left({\bf B}\right)\approx\frac{1}{2}\{\Gamma_{s_{\beta}}^{0}+\Gamma_{s_{\gamma}}^{0}\pm[(\Gamma_{s_{\beta}}^{0}-\Gamma_{s_{\gamma}}^{0})^{2}-4\omega_{B}^{2}]^{1/2}\}$,
$\omega_{B}\approx\mu_{B}g_{0}B$\tabularnewline
\hline 
\multirow{3}{*}{sv-E+D} & $\{V_{x}^{\mathrm{svR}},V_{y}^{\mathrm{svR}},V_{z}^{\mathrm{svR}}\}$ & $\Gamma_{s,i}^{\mathrm{sv-E+D}}$ = smallest eigenvalues of reduced
EVP Eq. \ref{eq:reduced_gep}.\tabularnewline
 & with & For Gr-hBN, define $\Omega^{s_{x}\theta s_{y}}=\varrho^{s_{x},\dagger}L^{e}\rho^{\theta s_{y}},$\tabularnewline
 & $V_{i}^{\mathrm{svR}}=\{U_{s_{i}}^{\mathrm{E+D},R},\rho^{\theta s_{i}}\}$ & $\Gamma_{s,x}^{\mathrm{sv-E+D}}\approx\frac{1}{2}\left(\Gamma_{s,x}^{\mathrm{E+D}}+\Gamma_{\theta}\right)-\frac{1}{2}[(\Gamma_{\theta}-\Gamma_{s,x}^{\mathrm{E+D}})^{2}-4|\Omega^{s_{x}\theta s_{y}}|^{2}]^{1/2}$\tabularnewline
\hline 
DD & $\left\{ U_{s}^{tR},L^{v_{j}}U_{s}^{tR}\right\} $ & $\lambda_{s}^{j}\approx[v_{F}^{-2}\tau_{p}^{-1}\Gamma_{s}]^{1/2}$\tabularnewline
\hline 
\multirow{3}{*}{ss-DD} & $\{V_{x}^{\mathrm{DDR}},V_{y}^{\mathrm{DDR}},V_{z}^{\mathrm{DDR}}\}$ & $\lambda_{s,i}^{j}$ = eigenvalues of reduced EVP Eq. \ref{eq:reduced_gep}.\tabularnewline
 & with & For Gr-hBN at ${\bf B}=0$, $\lambda_{s,x}^{x}=\lambda_{s,z}^{x}\approx[v_{F}^{-2}\tau_{p}^{-1}(\Gamma^{\mathrm{I}}\pm\Gamma^{\mathrm{II}})]^{1/2}$,\tabularnewline
 & $V_{i}^{\mathrm{DDR}}=\{U_{s_{i}}^{tR},L^{v_{j}}U_{s_{i}}^{tR}\}$ & $\Gamma^{\mathrm{I}}=\frac{\Gamma_{s,x}+\Gamma_{sz}-r\tau_{p}^{-1}}{2},\Gamma^{\mathrm{II}}=\frac{1}{2}[(\Gamma_{s,x}-\Gamma_{s,z})^{2}-\frac{r}{\tau_{p}}(2\Gamma_{s,x}+2\Gamma_{s,z}-\frac{r}{\tau_{p}})]^{1/2}$\tabularnewline
\hline 
\hline 
\end{tabular}\caption{Approximate formulas of complex spin relaxation rate $\Gamma_{s}$
and inverse spin diffusion length $\lambda_{s}$ derived from our
method (Eq. \ref{eq:reduced_gep}) without and with coupled dynamics.
Spin lifetime $\tau_{s}$=$1/\mathrm{Re}\Gamma_{s}$. Spin diffusion
length $l_{s}$=$1/\mathrm{Re}\lambda_{s}$. $V^{KR}$ - right Krylov
subspace (KS, see related text above Eq. \ref{eq:reduced_gep}). ``EY''
- Elliot-Yafet. ``E+D'' - EY+DP (Dyakonov-Perel). ``RR-2'' - RR
method with two-column KS. ``sv-E+D'' - E+D with spin-valley coupled
dynamics. ``DD'' - drift-diffusion. ``ss-DD'' - DD with spin-spin
coupled dynamics. $\rho^{s_{i}}$ is spin perturbative DM. $\varrho^{s_{i}}\propto s_{i}$
with $\varrho^{s_{i},\dagger}\rho^{s_{i}}$=1. $\tau_{p}$ is carrier
lifetime. $\Omega$ represents precession frequency. $\Gamma^{0}$
and $U^{0R}$ are zero-(${\bf B}$-)field quantities. $\rho^{\theta s_{i}}$
is valley-spin perturbative DM. $\Gamma_{\theta}$ is valley relaxation
rate. $U_{s}^{tR}$ is right eigenvector of $-L$ for $\tau_{s}$.
$v_{F}$ is Fermi velocity. $r$=$|v_{xz}^{\Gamma_{s}}|^{2}v_{F}^{-2}$
with $v_{xz}^{\Gamma_{s}}$=$U_{s_{x}}^{tL,\dagger}L^{v_{x}}U_{s_{z}}^{tR}$.
See derivations in Supplemental Material (SM)\citep{supplementalmaterial}.\label{tab:Approximate-formulae}}
\end{table*}

In this letter, we propose an \textit{ab initio} method of decay
processes of arbitrary observables and corresponding $l$ and $\tau$
without the above issues. Besides its predictive power and wide applicability,
the method has a key advantage: It enables clear \textit{ab initio}
descriptions of long $l$ and $\tau$, using approximate formulas
at different levels -- from the conventional ones like DD relation
to those beyond with coupled dynamics. Equivalently, the slow decay
of observables can be simulated via reduced master equations focusing
on a few coupled degrees of freedom such as spin, valley and current.
The method offers a transparent and powerful tool for understanding
diffusion and relaxation mechanisms, and promises to uncover new phenomena
driven by coupled dynamics across multiple degrees of freedom.

We first apply the Wigner transformation to non-linear DM ME.\citep{sekine2017quantum}
Suppose the total electronic Wigner distribution function corresponding
to the total DM is $\rho^{\mathrm{tot}}(t,{\bf R})$, with ${\bf R}$
real-space coordinate. Suppose $\rho^{\mathrm{tot}}$=$f$+$\rho$,
with $f$ Fermi-Dirac function and $\rho$ the non-equilibrium part.
Assuming $\rho$ is small, typical in device applications and measurements
of $\tau$ and $l$, we obtain linearized ME by linearizing the scattering
term:\citep{supplementalmaterial}
\begin{align}
\frac{d\rho_{\kappa}}{dt}+\Sigma_{j\kappa'}L_{\kappa\kappa'}^{v_{j}}\frac{d\rho_{\kappa'}}{dR_{j}}= & \Sigma_{\kappa'}L_{\kappa\kappa'}({\bf B})\rho_{\kappa'},\label{eq:linear}\\
L({\bf B})= & L^{e}({\bf B})+L^{C},\label{eq:L_eq_Le_LC}
\end{align}
where $\kappa$=$\{k,a,b\}$ is the combined index of k-point and
two band indices. $L^{v_{j}}$ relates to the diffusion and $(L^{v_{j}}\rho)_{kab}$$=$$(v_{jkac}\rho_{kcb}$$+$$\rho_{kac}v_{jkcb})/2$
with $v_{j}$ velocity matrix. $L^{e}\rho$ and $L^{C}\rho$ describe
the coherent and scattering dynamics respectively. $i\hbar(L^{e}\rho)_{kab}$=$[H_{k}^{e},\rho_{k}]_{ab}$=
$H_{kac}^{e}\rho_{kcb}$$-$$\rho_{kac}H_{kcb}^{e}$, where $H^{e}$
is field-dependent electronic Hamiltonian with a spin Zeeman term.
$L^{C}$ is determined by $f$ and the generalized scattering-rate
matrix. See details in Appendix A.\citep{xu2024ab,sekine2017quantum}

Solving Eq. \ref{eq:linear} is non-trivial in general. Here we consider
two important commonly-used\citep{wu2010spin} one-variable problems:
(i) Spatial homogeneous relaxation and (ii) 1D steady-state diffusion
along $x$. Their solutions are
\begin{align}
\rho_{\kappa}(X)= & \Sigma_{\mu}e^{-\eta_{\mu}^{X}X}U_{\kappa\mu}^{XR}\rho_{\mu}^{X},\label{eq:general_solution}
\end{align}
where $X$=$t,x$. $\mu$ is decay mode index. $\eta_{\mu}^{X}$ are
complex values and $\eta_{\mu}^{X}$=$\Gamma_{\mu}$ ($\lambda_{\mu}^{x}$)
for $X$=$t$ ($x$). $\mathrm{Re}\Gamma_{\mu}$ and $\mathrm{Re}\lambda_{\mu}^{x}$
are mode-resolved relaxation rate $\tau_{\mu}^{-1}$ and inverse diffusion
length $1/l_{\mu}^{x}$ respectively. $\mathrm{Im}\eta_{\mu}^{X}$
describes precession. $U_{\kappa\mu}^{XR}$ are right eigenvectors
of the eigenvalue problem (EVP):
\begin{align}
\Sigma_{\kappa'}A_{\kappa\kappa'}U_{\kappa'\mu}^{XR}= & \Sigma_{\kappa'}B_{\kappa\kappa'}U_{\kappa'\mu}^{XR}E_{\mu}^{X},\label{eq:gevp}
\end{align}
where $A$=$-L$, $B$=$I$, $E_{\mu}^{X}$=$\eta_{\mu}^{X}$ for
$X$=$t$, and $A$=$L^{v_{x}}$, $B$=$-L$, $E_{\mu}^{X}$=$1/\eta_{\mu}^{X}$
for $X$=$x$. Given the boundary condition $\rho$($X$=0$)$=$\rho^{\mathrm{pert}}$
with $\rho^{\mathrm{pert}}$ the perturbative DM, we obtain $\rho_{\mu}^{X}$=$\sum_{\kappa\kappa'}U_{\mu\kappa}^{XL,*}B_{\kappa\kappa'}\rho_{\kappa'}^{\mathrm{pert}}$
with $U_{\mu\kappa}^{XL}$ left eigenvector of Eq. \ref{eq:gevp}.
More discussions of the solutions of one- and two-variable decay problems
are given in Sec. SIII of SM\citep{supplementalmaterial}.

$\rho^{\mathrm{pert}}$ is chosen based on the physical problem: For
spin perturbation via spin Zeeman effect,\citep{xu2020spin} $\rho^{\mathrm{pert}}$=$\rho_{\kappa}^{s_{i}}$$\propto$$s_{i,\kappa}(\frac{\Delta f}{\Delta\epsilon})_{\kappa}$
with $s_{i}$ spin matrix. $\frac{\Delta f}{\Delta\epsilon}$ is $\frac{df}{d\epsilon}$
in degenerate subspace and $\frac{f_{ka}-f_{kb}}{\epsilon_{ka}-\epsilon_{kb}}$
otherwise. For valley-spin perturbation, $\rho^{\mathrm{pert}}$=$\rho_{kab}^{\theta s_{i}}$=$\theta_{k}\rho_{kab}^{s_{i}}$,
where $\theta_{k}$=$\pm1$ if $k$$\in$$\pm K$. Having the observable
operator $o_{\kappa}$ and defining $o_{\mu}^{X}$=$N_{k}^{-1}\sum_{\kappa}o_{\kappa}^{*}U_{\kappa\mu}^{XR}$,
the observable evolution is:\citep{supplementalmaterial}
\begin{align}
O(X)= & \mathrm{Re}(\Sigma_{\mu}\rho_{\mu}^{X}o_{\mu}^{X}e^{-\eta_{\mu}^{X}X}).\label{eq:OX}
\end{align}

Eq. \ref{eq:OX} accurately describes the observable decay for given
$\rho^{\mathrm{pert}}$. The observable dynamics consist of dynamics
of individual decay modes, highly simplifying the analysis compared
to previous real-time method\citep{xu2024spin}. For given $\rho^{\mathrm{pert}}$
and $o$, the relevance of a mode to $O(X)$ is measured by $\left|\rho_{\mu}^{X}o_{\mu}^{X}\right|$.
For slow decay of certain observables such as spin, typically only
a few modes are relevant, so that $O(X)$ is well described by eigenvalues
and eigenvectors of these ``relevant'' modes. If only one mode is
relevant, $\tau$ ($l^{x}$) of the observable is simply $\tau$ ($l^{x}$)
of this mode. However, with multiple non-degenerate relevant modes,
if we intend to describe the observable decay by a single $\tau$
or $l^{x}$, we need to define an effective $\tau$ ($l^{x}$) via
exponential-cosine fit of $O(t)$ ($O(x)$).

\begin{figure*}
\includegraphics[scale=0.65]{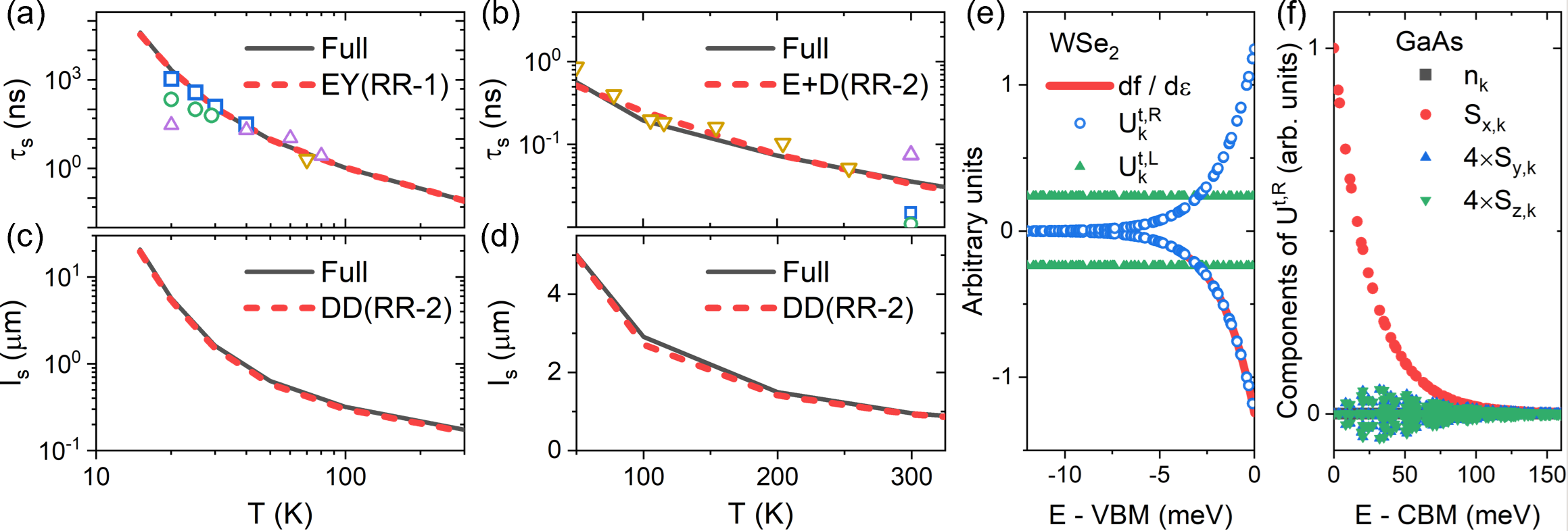}\caption{Theoretical results of spin relaxation and diffusion of holes of WSe$_{2}$
and electrons of GaAs. (a) and (b) are spin lifetimes $\tau_{s}$
of WSe$_{2}$ and GaAs respectively calculated by solving full EVP
(``Full'', Eq. \ref{eq:gevp}) and approximate formulas within the
RR method (Table \ref{tab:Approximate-formulae}). ``RR-$n$'' means
$n$-column KS are used. The points in subfigures are experimental
data.\citep{li2021valley,goryca2019detection,ersfeld2020unveiling,yan2017long,kimel2001room,bungay1997direct,dzhioev2004suppression}
(c) and (d) are spin diffusion lengths $l_{s}$ of WSe$_{2}$ and
GaAs respectively calculated by solving full EVP (``Full'', Eq.
\ref{eq:gevp}) and DD formula. (e) Eigenvectors $U_{k}^{t,L}$ and
$U_{k}^{t,R}$ corresponding to $\tau_{s,z}$ of WSe$_{2}$. (f) The
$k$-resolved carrier ($n_{k}$) and spin ($S_{i,k}$) components
of eigenvector $U^{t,R}$ corresponding to $\tau_{s,x}$ of GaAs.\label{fig:wse2_gaas}}
\end{figure*}

Low-power electronics often require slow decay of quantities like
spin for stable detection and manipulation of information. For slow
decay, it seems unnecessary to solve full EVPs (Eq. \ref{eq:gevp}),
but enough to obtain eigenvalues and eigenvectors of a few ``relevant''
modes using approximate methods, e.g., the Rayleigh-Ritz (RR) method\citep{macdonald1933successive}.
In this method, eigenvectors are linear combinations of trial vectors,
which span a Krylov subspace (KS)\citep{liesen2013krylov} here. For
a full EVP $AU^{R}=BU^{R}E$ (Eq. \ref{eq:gevp}), order-$n$ right
KS $V^{KR}$ consists of $A^{m-l}B^{l}V^{R}$ with $0\le l\le m\le n$,
where columns of $V^{R}$ are trial vectors. Similarly, left KS $V^{KL}$
consists of $A^{\dagger,m-l}B^{\dagger,l}V^{L}$ (see more details
in Appendix B). With $V^{KR\left(L\right)}$ and $M^{K}=V^{KL,\dagger}MV^{KR}$,
a reduced EVP is obtained
\begin{align}
A^{K}Y^{R}= & B^{K}Y^{R}E.\label{eq:reduced_gep}
\end{align}

Eigenvalues of Eq. \ref{eq:reduced_gep} are approximate eigenvalues
of the full EVP. Eigenvectors $U^{R(L)}$$\approx$$V^{KR(L)}Y^{R(L)}$
if enforcing $V^{KL,\dagger}B^{K}V^{KR}$=$I$.

Within the RR method, by selecting different KS (different $V^{R\left(L\right)}$
and order $n$), we can construct approximate formulas of long $\tau$
and $l$ at different levels. In Table \ref{tab:Approximate-formulae},
we present various formulas of $\tau_{s}$ and $l_{s}$: \textbf{(i)
EY formula} of $\tau_{s}$, where spin relaxation is due to spin-flip
scattering.\citep{vzutic2004spintronics} \textbf{(ii) Generalized
E+D formula} of $\tau_{s}$: It considers both EY and DP mechanisms.
In strong scattering limit $\tau_{p}^{-1}\gg2(\overline{\Omega_{\perp i}^{2}})^{1/2}$
and assuming $\tau_{p}^{-1}\gg\text{\ensuremath{\Gamma_{s}^{\mathrm{EY}}}}$,
it reduces to $\Gamma_{s,i}^{\mathrm{E+D}}\approx\Gamma_{s,i}^{\mathrm{EY}}+\Gamma_{s,i}^{\mathrm{DP}}$
with $\Gamma_{s,i}^{\mathrm{DP}}=\tau_{p}\overline{\Omega_{\perp i}^{2}}$
being actually the DP relation.\citep{vzutic2004spintronics} \textbf{(iii)
sv-E+D formula, i.e., E+D formula with spin-valley (sv) coupled dynamics},
where dynamics of spins and valley spins are coupled. \textbf{(iv)
DD formula} of $l_{s}$: Let $v_{F}^{2}\tau_{p}$ the diffusion coefficient
$D_{s}^{j}$, we have $\lambda_{s}^{j}\approx[(D_{s}^{j})^{-1}\Gamma_{s}]^{1/2}$,
which is indeed the well-know DD relation.\citep{vzutic2004spintronics,wu2010spin}
The formula also applies to spin diffusion at a transverse ${\bf B}$,
where spins precess about ${\bf B}$. Then $\Gamma_{s}$ is complex
and $\Gamma_{s}\approx\tau_{s}^{-1}+i\omega_{B}$ with $\omega_{B}=\mu_{B}g_{0}B$,
so that $l_{s}=l_{s}^{B=0}(\frac{1}{2}+\frac{1}{2}\sqrt{1+\tau_{s}^{2}\omega_{B}^{2}})^{-1/2}$
with $l_{s}^{B=0}$ the zero-field value. \textbf{(v) ss-DD formula,
i.e., DD formula with spin-spin (ss) coupled dynamics}, where dynamics
of spins along different directions are coupled.

\begin{figure}
\includegraphics[scale=0.7]{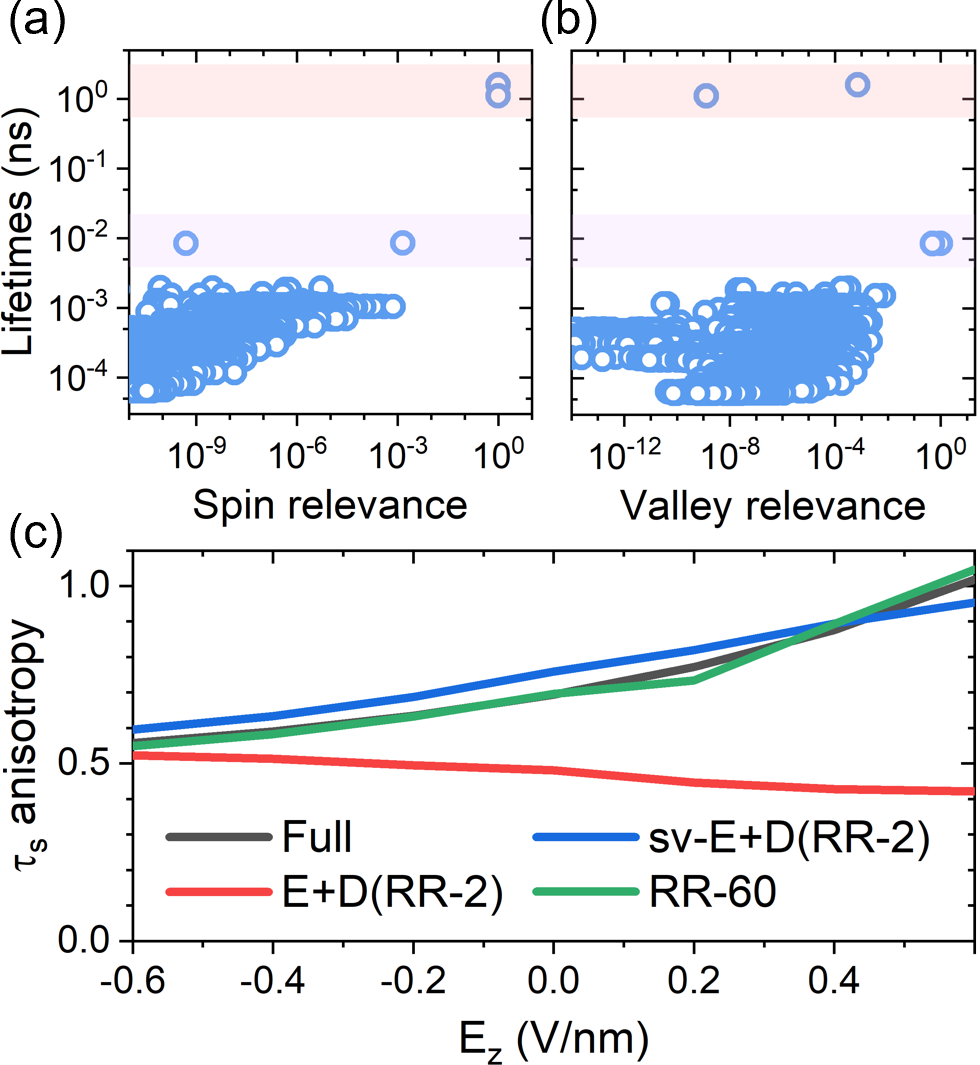}\caption{Lifetimes and $\tau_{s}$ anisotropy of graphene-hBN at 300 K and
$E_{F}$=0.1 eV. (a) and (b) show lifetimes of all decay modes with
their spin relevance and valley relevance respectively (Appendix C).
(c) Calculated $\tau_{s}$ anisotropy ratios as a function of electric
field $E_{z}$. The differences between results by E+D and sv-E+D
formulas indicate the importance of the sv coupled dynamics.\label{fig:gr-bn_lifetime}}
\end{figure}

\begin{figure*}
\includegraphics[scale=0.65]{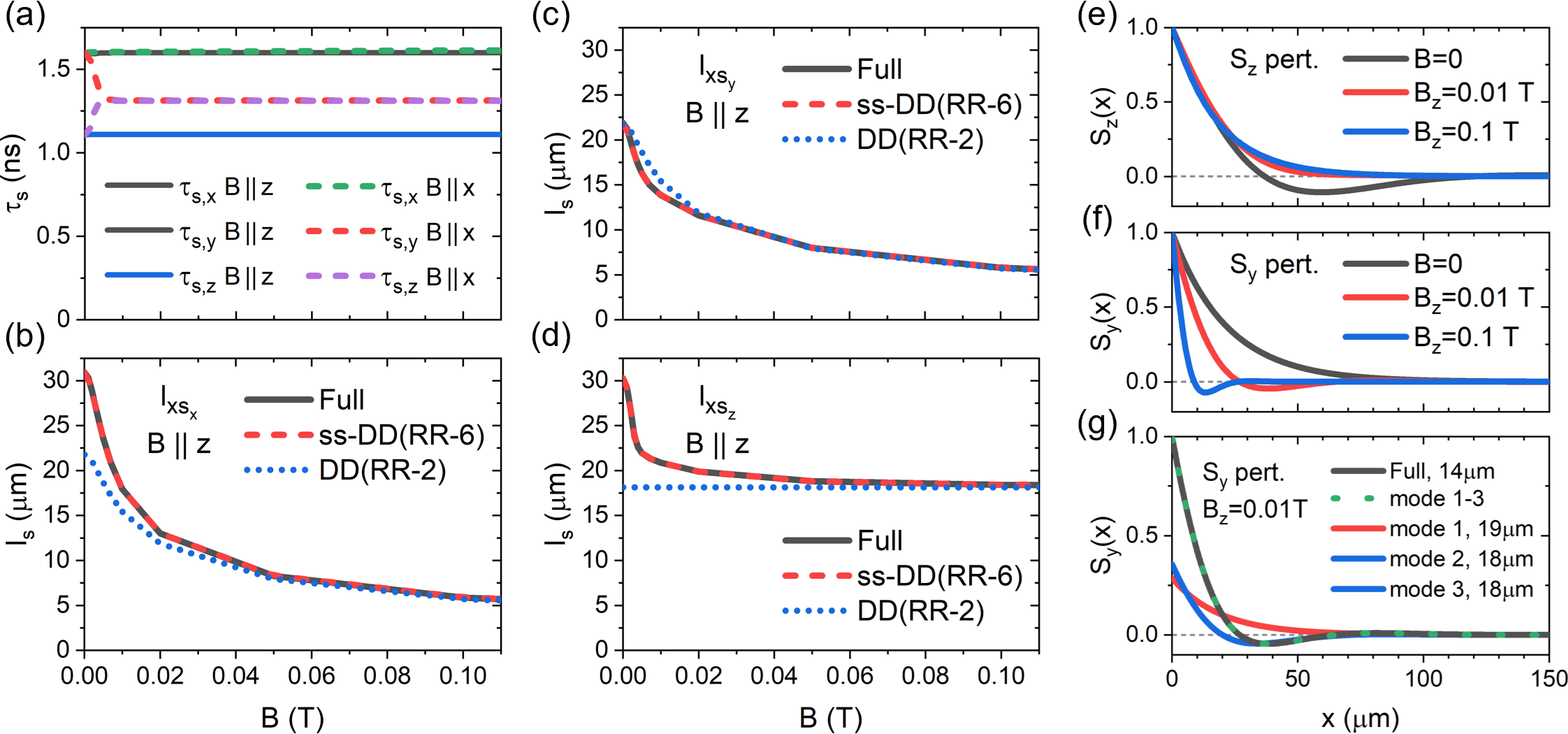}\caption{Magnetic-field ${\bf B}$ dependence of spin relaxation and diffusion
in graphene-hBN. (a) Calculated $\tau_{s}$ at ${\bf B}$ along different
directions. (b), (c) and (d) are calculated $l_{xs_{x}}$, $l_{xs_{y}}$
and $l_{xs_{z}}$ respectively as a function of $B$ with ${\bf B}||{\bf z}$.
The differences between results by DD and ss-DD formulas indicate
the importance of the ss coupled dynamics. (e) and (f) are 1D steady-state
solutions of $S_{z}\left(x\right)$ and $S_{y}\left(x\right)$ respectively
at different ${\bf B}$ along $z$. (g) $S_{y}\left(x\right)$ at
$B=$0.01 T due to different decay modes.\label{fig:gr-bn_diffusion_Bext}}
\end{figure*}

Moreover, with small KS, we obtain a reduced ME of a small vector
$\mathbb{S}$=$V^{KL,\dagger}B^{K}\rho$: $d\mathbb{S}/dt$$+$$\Sigma_{j}\widetilde{L}^{v_{j}}(d\mathbb{S}/dR_{j})$=$\widetilde{L}\mathbb{S}$,
with $\widetilde{M}$=$(I^{K})^{-1}M^{K}$ and $I^{K}$=$V^{KL,\dagger}V^{KR}$.
If $V^{KR(L)}$ is closely related to spins or spin currents, the
equation approximately describes spin decay processes.

We next apply our method to simulate $\tau_{s}$ and $l_{s}$ due
to spin-orbit coupling (SOC) and e-ph scattering in a few spintronic
materials. See technical details in Sec. SIV of SM\citep{supplementalmaterial}.
In Fig. \ref{fig:wse2_gaas}(a)-(d), we show theoretical $\tau_{s}$
and $l_{s}$ of monolayer WSe$_{2}$ holes and GaAs electrons. Our
calculated $\tau_{s}$ are in agreement with experiments shown here
and previous \textit{ab initio} results\citep{xu2021ab,park2022predicting}.
Results from solving full EVP (Eq. \ref{eq:gevp}) match perfectly
those from approximate formulas within the RR method (Table \ref{tab:Approximate-formulae}),
validating the corresponding relaxation mechanisms. The eigenvector
analysis for the full $-L$ matrix in Fig. \ref{fig:wse2_gaas}(e)-(f)
reveals: (i) For WSe$_{2}$, $U^{tR}$ matches $df/d\epsilon$ in
one valley and $-df/d\epsilon$ in the other, confirming $U^{tR}=\rho^{s_{z}}$
(as $s_{z}=\pm1$ in two valleys). Similarly, $U^{tL}$ matches $\varrho^{s_{z}}\propto s_{z}$.
Such observations validate our EY formula. (ii) For $\tau_{s,x}$
of GaAs, $U^{tR}$ is dominated by $S_{x}$ component but with small
$S_{y}$ and $S_{z}$ components. The $k$ distribution of $S_{x}$
component fits $-df/d\epsilon$ function, suggesting that $U^{tR}$
is dominated by $\rho^{s_{x}}$. Further analysis shows that $U^{tR}$
is similar to the approximate right eigenvector of E+D formula within
the RR method - $\approx(1+\tau_{p}L)\rho^{s_{i}}$. These analysis
indicate the accuracy of our E+D formula.

In Fig. \ref{fig:gr-bn_lifetime}(a)-(b), we analyse spin and valley
relevances of all decay modes near Fermi level $E_{F}$ of graphene-hBN
heterostructure (Gr-hBN) and find: Three modes are relevant to spin
relaxation - two for $\tau_{s,x}$ and $\tau_{s,y}$ ($\approx$1.6
ns) and one for $\tau_{s,z}$ ($\approx$1.1 ns). Four modes are relevant
to valley relaxation ($\tau\approx$8.5 ps) - one for $\theta$ and
three for $\theta s_{i}$. Other modes have shorter $\tau$ values
(10 fs to 2 ps) and their $\tau^{-1}$ magnitudes match the diagonal
elements of $-L^{C}$, denoted as $-L_{kab}^{C,\mathrm{d}}$. $-L_{kab}^{C,\mathrm{d}}$
are actually state-resolved carrier relaxation rates $\tau_{p,ka}^{-1}$
when $a=b$. Thus, the large variance of short $\tau$ values reflects
the wide distribution of $\tau_{p}$ across k-points or state energies.

We then show calculated $\tau_{s}$ anisotropy ratios ($\tau_{s,z}/\tau_{s,x}$)
of Gr-hBN as a function of out-of-plane electric field\citep{xu2021giant,xu2023substrate}
($E_{z}$) in Fig. \ref{fig:gr-bn_lifetime}(c). $\tau_{s,z}/\tau_{s,x}$
by solving full EVP (black line) is found 0.7 at $E_{z}=0$ and increasing
with $E_{z}$, consistent with experiments\citep{guimaraes2014controlling}
and real-time simulations\citep{habib2022electric}. Our E+D formula
(red line) incorrectly predicts $\tau_{s,z}/\tau_{s,x}\approx$0.5,
as the conventional DP relation. The issue is resolved by sv-E+D formula
(blue line), accounting for the sv coupled dynamics. We find that
$\tau_{s,z}$ is not affected by such coupled dynamics but $\tau_{s,x}$
is shorten by $S_{x}$-$\theta S_{y}$ coupled dynamics. Therefore,
$\tau_{s,x}$ can be simulated using the RR method with a two-column
KS, where $V^{KR}=\{U_{s_{x}}^{\mathrm{E+D},R},\rho^{\theta s_{y}}\}$,
which leads to an analytical formula of spin relaxation rate - $\Gamma_{s,x}^{\mathrm{sv-E+D}}$
present in Table \ref{tab:Approximate-formulae}. Assuming $\Gamma_{\theta}\gg|\Omega^{s_{x}\theta s_{y}}|$,
the formula reduces to $\Gamma_{s,x}^{\mathrm{sv-E+D}}$$\approx$$\Gamma_{s,x}^{\mathrm{E+D}}$+$\tau_{\theta}|\Omega^{s_{x}\theta s_{y}}|^{2}$,
indicating that $S_{x}$ relaxation is enhanced by an intervalley
term $\tau_{\theta}|\Omega^{s_{x}\theta s_{y}}|^{2}$ besides the
usual EY+DP contribution $\Gamma_{s,x}^{\mathrm{E+D}}$. In Gr-hBN,
the coherent term $L^{e}\rho$ describes Larmor precession about SOC
fields ${\bf B}^{\mathrm{soc}}$ with frequency ${\bf \Omega}^{\mathrm{soc}}$=$(e/m_{e}){\bf B}^{\mathrm{soc}}$.
Thus, $\Omega^{s_{x}\theta s_{y}}$=$\varrho^{s_{x},\dagger}L^{e}\rho^{\theta s_{y}}$$\approx$$\langle\widehat{x}$$\cdot$$({\bf \Omega}_{k}^{\mathrm{soc}}$$\times$$\theta_{k}\widehat{y})\text{\ensuremath{\rangle}}$,
with $\langle F\rangle$ Fermi-surface averaged $F$. Then, $\Omega^{s_{x}\theta s_{y}}$$\approx$$\langle\Omega_{z}^{\mathrm{soc}}\rangle_{K'}$$-$$\langle\Omega_{z}^{\mathrm{soc}}\rangle_{K}$,
where $\langle\rangle_{Q}$ means averaging within valley $Q$. Therefore,
with $|\Omega^{s_{x}\theta s_{y}}|$ reflecting the fluctuation of
$\Omega_{z}^{\mathrm{soc}}$ between valleys (see spin texture in
Fig. S2 of SM\citep{supplementalmaterial}) and $\tau_{\theta}$ intervalley
lifetime, $\tau_{\theta}|\Omega^{s_{x}\theta s_{y}}|^{2}$ represents
an intervalley DP contribution to spin relaxation.\citep{habib2022electric,cummings2017giantspinlifetime}
The results within the RR method can be further improved by using
high-order KS with more columns, e.g., 60 (``RR-60'', green line).

Magnetic field (${\bf B}$) is integral to tuning material properties
and measuring $\tau_{s}$ and $l_{s}$. In Fig. \ref{fig:gr-bn_diffusion_Bext}(a),
we study ${\bf B}$-field effect on $\tau_{s}$. For ${\bf B}||{\bf z}$,
$\tau_{s}$ are unaffected by ${\bf B}$. However, for ${\bf B}||x$,
spin relaxations along $y$ and $z$ mix, and both $\tau_{s,y}$ and
$\tau_{s,z}$ converge to an intermediate value as $B$ increases.
These results align with the approximate ``RR-2'' formula in Table
\ref{tab:Approximate-formulae}, derived from the RR method with two-column
KS accounting for $S_{\beta}$-$S_{\gamma}$ coupled dynamics ($\beta\neq\gamma$
and $\beta,\gamma\perp{\bf B}$).

We further examine spin diffusion in Gr-hBN at different ${\bf B}$
along $z$ (Fig. \ref{fig:gr-bn_diffusion_Bext}(b)-(g)). Our predicted
$l_{s}$ at ${\bf B}$=0 are 21-31 $\mu$m, at the upper bound of
the experimental range (a few to 31 $\mu$m)\citep{avsar2020colloquium}.
While DD formula predicts accurate $l_{s}$ at high ${\bf B}$, it
fails at low ${\bf B}$. The issue is removed by ss-DD formula, indicating
the importance of the ss coupled dynamics. From Fig. \ref{fig:gr-bn_diffusion_Bext}(e)-(f),
we find: At $B\ge$0.01 T, spins precess about ${\bf B}$ as expected;
However, at ${\bf B}=0$, spin precession is observed for $S_{z}$
perturbation but not for $S_{y}$. The phenomenon of zero-field spatial
spin precession is absent in DD formula but captured by ss-DD formula.
Such phenomenon was reported in previous model studies of quantum
wells\citep{weng2004spin} and observed in experiments\citep{crooker2005imaging}.
Here we provide a clear \textit{ab initio} description of it. For
Gr-hBN, the $S_{x}$-$S_{z}$ coupled dynamics yields $\lambda_{s,z}^{x}$=$\lambda_{s,x}^{x}$$\approx$$[(\overline{v_{xs}^{2}})^{-1}\tau_{p}^{-1}(\Gamma^{\mathrm{I}}\pm\Gamma^{\mathrm{II}})]^{1/2}$
(Table \ref{tab:Approximate-formulae}). It reduces to DD formula
if $|v_{xz}^{\Gamma_{s}}|$=$|v_{zx}^{\Gamma_{s}}|$=0. For Gr-hBN,
$|v_{xz}^{\Gamma_{s}}|$$\approx$$|v_{zx}^{\Gamma_{s}}|$$\approx$0.01
is however large, invalidating DD formula.

The low-field differences between DD and ss-DD formulas in Gr-hBN
can be understood via the approximate ME: $\frac{d\rho_{kab}}{dx}$=$v_{xk}^{-1}\{$$\frac{-i}{\hbar}[H_{k}^{\sigma},\rho_{k}]_{ab}$+$(L^{C}\rho)_{kab}\}$,
valid in many non-magnetic systems with weak SOC. $H_{kab}^{\sigma}$$\propto$$({\bf B}$+${\bf B}_{k}^{\mathrm{soc}})$$\cdot$${\bf \sigma}_{ab}$
with ${\bf \sigma}$ Pauli operator. Spatially, spins precess about
the ``spatial fields'' ${\bf B}_{k}^{v_{x}}$=$v_{xk}^{-1}$$({\bf B}$+${\bf B}_{k}^{\mathrm{soc}})$.
At low ${\bf B}$, for Rashba SOC (${\bf B}_{k}^{\mathrm{soc}}$$\propto$$(k_{y},-k_{x},0)$),
$\langle{\bf B}^{v_{x}}\rangle$ is finite along $y$, causing global
precession that couples $S_{x}$ and $S_{z}$ dynamics. Such coupling
is missing in DD formula and significantly affects diffusion of $S_{x}$
and $S_{z}$. For GaAs and WSe$_{2}$, $\langle{\bf B}^{v_{x}}\rangle$
is either $\approx$0 or along $z$, resulting in no zero-field precession
for GaAs and $S_{z}$ of WSe$_{2}$, matching our $S(x)$ results.
At high ${\bf B}$, precession is however caused by ${\bf B}$, so
that with complex ${\bf B}$-dependent $\Gamma_{s}$, DD formula still
works.

Furthermore, Fig. \ref{fig:gr-bn_diffusion_Bext}(g) shows $S_{y}$
diffusion at low ${\bf B}$ due to different modes. Notably, the effective
$l_{s}$ (14 $\mu$m) from exponential-cosine fit of $S_{y}(x)$,
is shorter than $l_{s}$ of each mode ($\ge$18 $\mu$m). This underscores
the necessity of spatial evolution simulations via Eq. \ref{eq:OX}
when there are multiple non-degenerate modes with complex eigenvalues.

Similar to Gr-hBN, our results of GaN (Fig. S1 of SM\citep{supplementalmaterial})
show strong $B_{z}$-dependence of $l_{s}^{x}$ and zero-field precession
for $S_{z}$. These are captured by ss-DD formula but not DD formula,
confirming the importance of ss coupled dynamics. Our $l_{s}$ results
of Pt and antiferromagnetic RuO$_{2}$ are in agreement with experiments
(Sec. SI of SM\citep{supplementalmaterial}), further verifying the
method's reliability.

\section*{Acknowledgments}

J.X. thanks Ravishankar Sundararaman, Chong Wang and Zhengzheng Chen
for helpful discussions. This work is supported by National Natural
Science Foundation of China (Grant No. 12304214), Fundamental Research
Funds for Central Universities (Grant No. JZ2023HGPA0291). This research
used resources of the HPC Platform of Hefei University of Technology.


%

\section*{End Mater}

\subsection*{Appendix A: The linearized DM ME}

We apply the Wigner transformation to the non-linear density-matrix
(DM) master equation (ME) and include the diffusion term. We then
obtain\citep{sekine2017quantum}
\begin{align}
 & \frac{d\rho_{\kappa}^{\mathrm{tot}}\left(t,{\bf R}\right)}{dt}+\frac{1}{2}\left\{ {\bf v}\cdot\frac{d\rho^{\mathrm{tot}}\left(t,{\bf R}\right)}{d{\bf R}}\right\} _{\kappa}\nonumber \\
= & -\frac{i}{\hbar}\left[H^{e}\left({\bf B}\right),\rho^{\mathrm{tot}}\right]_{\kappa}+C_{\kappa}\left[\rho^{\mathrm{tot}}\right],\label{eq:nonlinear}
\end{align}
where $\frac{1}{2}\left\{ {\bf v}\cdot\nabla_{{\bf R}}\rho^{\mathrm{tot}}\right\} $
is the diffusion term with ${\bf v}$ the vector of velocity matrices.
$\left\{ {\bf a}\cdot{\bf b}\right\} $$=$${\bf a}\cdot{\bf b}$$+$${\bf b}\cdot{\bf a}$.
The right-hand-side terms are the coherent and scattering terms respectively.
$H^{e}({\bf B})$$=H^{e}(0)$$+$$H^{\mathrm{sZ}}({\bf B})$, with
$H^{\mathrm{sZ}}({\bf B})=\mu_{B}g_{0}{\bf B}\cdot{\bf s}$ being
spin Zeeman Hamiltonian. $\left[H,\rho\right]=H\rho-\rho H$. Within
Born-Markov approximation and neglect the renormalization part, the
scattering term $C\left[\rho^{\mathrm{tot}}\right]$ reads\citep{xu2024spin,xu2023ab}
\begin{align}
C_{kab}\left[\rho^{\mathrm{tot}}\right]= & \frac{1}{2}\sum_{ck'de}\left[\begin{array}{c}
\left(I-\rho^{\mathrm{tot}}\right)_{kac}P_{kcb,k'de}\rho_{k'de}^{\mathrm{tot}}\\
-\left(I-\rho^{\mathrm{tot}}\right)_{k'de}P_{k'de,kac}^{*}\rho_{kcb}^{\mathrm{tot}}
\end{array}\right]\nonumber \\
 & +H.C.,
\end{align}
where $P$ is the generalized scattering-rate matrix. Suppose $P^{\mathrm{e-ph}}$,
$P^{\mathrm{e-i}}$ and $P^{\mathrm{e-e}}$ are $P$ of the e-ph,
electron-impurity (e-i) and electron-electron (e-e) scatterings. Assuming
different scattering processes are independent, the total $P$ is
$P=P^{\mathrm{e-ph}}+P^{\mathrm{e-i}}+P^{\mathrm{e-e}}$. $P^{\mathrm{e-ph}}$
and $P^{\mathrm{e-i}}$ are independent from $\rho^{\mathrm{tot}}\left(t,{\bf R}\right)$
and computed from first-principles energies and corresponding scattering
matrix elements. $P^{\mathrm{e-e}}$ is more complicated, since it
depends on not only energies and scattering matrix elements but also
$\rho^{\mathrm{tot}}\left(t,{\bf R}\right)$.\citep{xu2023ab}

Suppose $\rho^{\mathrm{tot}}$=$f$+$\rho$ and assume $\rho$ is
small. By linearizing the scattering term, the nonlinear ME (Eq. \ref{eq:nonlinear})
can be rewritten as
\begin{align}
\frac{d\rho_{\kappa}}{dt}+\sum_{j\kappa'}L_{\kappa\kappa'}^{v_{j}}\frac{d\rho_{\kappa'}}{dR_{j}}= & \sum_{\kappa'}L_{\kappa\kappa'}({\bf B})\rho_{\kappa'}+\sum_{\kappa'}L_{\kappa\kappa'}^{e}({\bf B})f_{\kappa'},\label{eq:linear_withf}
\end{align}
where $L({\bf B})=L^{e}({\bf B})+L^{C}$. In the above equation, we
have considered $L^{C}f=0$, reflecting the fact that there is no
scattering at equilibrium. $L^{v_{j}}$ and $L^{e}$ are
\begin{align}
L_{kab,k'cd}^{v_{j}}= & \frac{1}{2}\left(v_{j,kac}\delta_{bd}+\delta_{ac}v_{j,kdb}\right)\delta_{kk'},\\
L_{kab,k'cd}^{e}= & \frac{-i}{\hbar}\left(H_{kac}^{e}\delta_{bd}-\delta_{ac}H_{kdb}^{e}\right)\delta_{kk'}.
\end{align}

At ${\bf B}=0$, $H_{kab}^{e}=\epsilon_{ka}\delta_{ab}$, we can write
$L^{e}$ as $L^{\epsilon}$,
\begin{align}
L_{kab,k'cd}^{\epsilon}= & \frac{-i}{\hbar}\left(\epsilon_{ka}-\epsilon_{kb}\right)\delta_{ac}\delta_{bd}\delta_{kk'}.
\end{align}

At ${\bf B}\neq0$, by choosing eigenstates of $H^{e}\left({\bf B}\right)$
as basis functions, $H^{e}\left({\bf B}\right)$ can be written as
$H^{e}\left({\bf B}\right)=\epsilon_{ka}\left({\bf B}\right)\delta_{ab}$,
so that the $L^{e}$ again has the above form.

In this work, the basis are always eigenstates of $H^{e}\left({\bf B}\right)$,
leading to
\begin{align}
\sum_{\kappa'}L_{\kappa\kappa'}^{e}({\bf B})f_{\kappa'}= & 0,
\end{align}
so that Eq. \ref{eq:linear_withf} becomes Eq. \ref{eq:linear}. If
${\bf B}\neq0$ and the basis are chosen as eigenstates of zero-field
Hamiltonian $H^{e}\left(0\right)$, the constant term $L^{e}\left({\bf B}\right)f$
is then non-zero in general. This term can be eliminated, e.g., by
variable substitution or redefining the equilibrium DM. Nevertheless,
the constant term is found unimportant and does not affect $\tau$
and $l$ at all in this study. Numerically, we find that calculated
$\tau$ and $l$ obtained with different types of basis are identical.

For the e-ph scatting, $L^{C}$ reads
\begin{align}
L_{kab,k'cd}^{C}= & \overline{L}_{kab,k'cd}^{C}+\overline{L}_{kba,k'dc}^{C,*},\\
\overline{L}_{kab,k'cd}^{C}= & \frac{1}{2N_{k}}\left[\left(I-f\right)_{ka}P_{kab,k'cd}+P_{k'cd,kab}^{*}f_{kb}\right]\nonumber \\
 & -\frac{1}{2N_{k}}\delta_{kk'}\sum_{k''e}\delta_{ac}P_{kdb,k''ee}f_{k''e}\nonumber \\
 & -\frac{1}{2N_{k}}\delta_{kk'}\sum_{k''e}\left(I-f\right)_{k''e}P_{k''ee,kac}^{*}\delta_{bd}.
\end{align}

For the e-i scattering, the form of $L^{C}$ is the same. For the
e-e scattering, although the formula of $L^{C}$ is more complicated
due to the $\rho^{\mathrm{tot}}\left(t,{\bf R}\right)$ dependence
of $P^{\mathrm{e-e}}$, but its derivation is straightforward.

In this work, the electric field along periodic direction and the
laser are not considered but can be done using covariant derivative.\citep{xu2024ab,sekine2017quantum}
Therefore, our framework may be generalized to simulate other transport
properties and laser-related dynamical or transport phenomena. These
are beyond our current scope and can be done in the future.

\subsection*{Appendix B: The Rayleigh-Ritz (RR) method and Krylov subspaces (KS)}

Within the RR method, for the generalized EVP
\begin{align}
AU^{R}= & BU^{R}E,\label{eq:original_gep}
\end{align}
alternative order-$n$ right and left Krylov subspaces (KS) are
\begin{align}
V^{KR,o(n)}= & \{AV^{KR,o(n-1)},BV^{KR,o(n-1)}\},\label{eq:VKR}\\
V^{KR,o(0)}= & \{V^{R}\},\\
V^{KL,o(n)}= & \{A^{\dagger}V^{KL,o(n-1)},B^{\dagger}V^{KL,o(n-1)}\},\label{eq:VKL}\\
V^{KL,o(0)}= & \{V^{L}\},
\end{align}
where $V^{KR/L}$ is right/left KS. ``$o(n)$'' means order-$n$.
Each column of $V^{R\left(L\right)}$ is a trial vector. Note that
if a column of $V^{KR/L}$ is a linear combination of other columns,
this column should be removed from $V^{KR/L}$. Alternatively, the
columns of $V^{R}$ and $V^{L}$ can be chosen as different $\rho^{\mathrm{pert}}$
and $o$ respectively. When $B=I$, the order-$n$ KS is
\begin{align}
V^{KR,o(n)}= & \left\{ V^{R},AV^{R},A^{2}V^{R},......,A^{n}V^{R}\right\} ,\\
V^{KL,o(n)}= & \left\{ V^{L},A^{\dagger}V^{L},\left(A^{\dagger}\right)^{2}V^{L},......,\left(A^{\dagger}\right)^{n}V^{L}\right\} .
\end{align}

With $V^{KR\left(L\right)}$, a new generalized EVP (Eq. \ref{eq:reduced_gep})
is obtained. The eigenvalues of Eq. \ref{eq:reduced_gep} are approximate
eigenvalues of the original full EVP (Eq. \ref{eq:original_gep}).
Therefore, approximate formulas of the eigenvalue-related quantities
(such as $l$ and $\tau$) can be constructed by selecting different
$V^{KR(L)}$, i.e., different $V^{R(L)}$, $n$ (order). Suppose $V^{R}=\left\{ \rho^{\mathrm{pert}}\right\} $
and $V^{L}=\left\{ o\right\} $. If $n$ are sufficiently large, the
accuracy of the RR method (Eq. \ref{eq:VKR} and \ref{eq:VKL}) should
be the same as solving the observable evolution corresponding to operator
$o$ from the full linearized ME (Eq. 1 in the main text) with $\rho\left(X=0\right)=\rho^{\mathrm{pert}}$.

\subsection*{Appendix C: Spin relevance and valley relevance}

We first define normalized observable operator $\widehat{o}\propto o$
with $\sum_{\kappa}\widehat{o}_{\kappa}^{*}\widehat{o}_{\kappa}=1$.
Assuming $\rho^{\mathrm{pert}}$ are all normalized, i.e., $\sum_{\kappa}\rho_{\kappa}^{\mathrm{pert},*}\rho_{\kappa}^{\mathrm{pert}}=1$,
we define spin and valley relevance to a given time-decay mode $\nu$
- $\mathscr{R}_{\nu}^{s}$ and $\mathscr{R}_{\nu}^{\theta}$ as
\begin{align}
\mathscr{R}_{\nu}^{s}= & \sqrt{\sum_{i}\left(\sum_{\kappa}\widehat{s}_{i,\kappa}^{*}U_{\kappa\nu}^{tR}\sum_{\kappa'}U_{\kappa'\nu}^{tL,*}B\rho_{\kappa'}^{s_{i}}\right)^{2}},
\end{align}
\begin{align}
\mathscr{R}_{\nu}^{\theta}= & \sqrt{\left(\mathscr{R}_{\nu}^{\theta\mathrm{I}}\right)^{2}+\left(\mathscr{R}_{\nu}^{\theta\mathrm{II}}\right)^{2}},
\end{align}
\begin{align}
\mathscr{R}_{\nu}^{\theta\mathrm{I}}= & \sum_{\kappa}\widehat{\theta}_{\kappa}^{*}U_{\kappa\nu}^{tR}\sum_{\kappa'}U_{\kappa'\nu}^{tL,*}B\rho_{\kappa'}^{\theta},
\end{align}
\begin{align}
\mathscr{R}_{\nu}^{\theta\mathrm{II}}= & \sqrt{\sum_{i}\left(\sum_{\kappa}\left(\widehat{\theta}^{s_{i}}\right)_{\kappa}^{*}U_{\kappa\nu}^{tR}\sum_{\kappa'}U_{\kappa'\nu}^{tL,*}B\rho_{\kappa'}^{\theta s_{i}}\right)^{2}},
\end{align}
where $\widehat{\theta}^{s_{i}}$ is the normalized $\theta s_{i}$
and $\rho_{kab}^{\theta}\propto\theta_{k}\frac{df_{ka}}{d\epsilon}\delta_{ab}$.
\end{document}